\documentclass[aps,twocolumn,floatfix,showpacs]{revtex4}
\usepackage{graphicx,bm}
\usepackage{amsmath}
\begin{document}
\title{Pseudopotential in resonant regimes}
\author{Ludovic Pricoupenko}
\affiliation
{Laboratoire de Physique Th\'{e}orique de la Mati\`{e}re Condens\'{e}e, 
Universit\'{e} Pierre et Marie Curie, 4 place Jussieu,
75252 Paris Cedex 05, France.}
\date{\today}
\begin{abstract}
The  zero-range potential approach is extended  for the description of
situations where two-body scattering is  resonant in arbitrary partial
waves.  The formalism generalizes the  Fermi pseudopotential which can
be used  only for $s$-wave broad  resonances. In a given  channel, the
interaction is described either in terms of a contact condition on the
wave function or  with a family  of pseudopotentials.  We show that it is
necessary to introduce a regularized scalar product for wave functions
obtained in  the zero-range potential  formalism (except for the Fermi
pseudopotential).   This metrics shows that  the geometry of these
Hilbert spaces depends crucially on the interaction.
\end{abstract}
\maketitle
\section{Introduction}

\subsection{Zero-range approach in ultracold physics}

The roots of the zero-range potential approach lie in the works on the
neutron-proton interaction   in  nuclear  physics.   First,  Bethe and
Peierls  \cite{Bethe}  showed that  it  was possible  to get  numerous
scattering  properties on deuteron   by  replacing the  neutron-proton
force with a  suitable boundary condition on  the wave function in the
$s$-wave channel. One year after,   Fermi \cite{Fermi} introduced  the
idea of a zero-range effective potential allowing a Born approximation
in the computation   of the scattering cross section  between a slow
neutron and a proton.    Equivalence between these two  approaches was
developed further  by  Breit  \cite{Breit}  and  led to  the  final
formulation of the  so-called Fermi pseudopotential \cite{Blatt}.  The
fact that  the range of the nuclear  force (${\simeq2 × 10^{-15}}$~m) is much
less than the $s$-wave neutron-proton  triplet scattering length  (${\simeq
4.3 × 10^{-15}}$~m)   was a  key ingredient  for   the  success of  this
approach.

More than  half   a  century after  these   pioneer  works, zero-range
pseudopotentials    appear  as reference  tools  in    the physics  of
ultracold   atoms.   There  are   several reasons   for this  renewed
interest.  First, the mean  interparticle distance in ultracold gases
is very large as compared to  the range of interatomic forces (denoted
by  $R$), so  that  two-body  collisions  occur at energies  very much
smaller than the characteristic high-energy scale  $\hbar^2 /mR^2$ ($m$ is
the particle mass). Consequently, the full two-body potential contains
useless informations  (deep  bound states, hard-core  repulsion \dots)
for the study  of the ultracold dilute  gases.  In this  respect, the
pseudopotentials   which  by   construction   reproduce the    correct
scattering phase shift at small energy,  contain the minimal number of
parameters for modeling two-body collisions.  A second key property is
that for  processes of   vanishly small energies,  a large    class of
pseudopotentials  give  the correct  results  at the  first order Born
approximation while this approximate  treatment is not possible  using
the ``real'' two-body  potential.  Hence, mean-field  schemes like the
Gross-Pitaevskii and the Bogoliubov approaches
\cite{Cohen,YvanLesHouches} for the Bose gas or BCS \cite{BruunBCS} for 
the dilute two-component Fermi gas can  be implemented using the Fermi
pseudopotential.  This  idea  has  been generalized by   introducing a
family of pseudopotentials  equivalent  to the Fermi  pseudopotential:
the $\lambda$-potentials  \cite{Vlambda}.  Varying  the  free parameter  $\lambda$
does not change the exact treatment of the two-body problem, while the
Born approximation can be adjusted to  the exact solution for a finite
collisional energy.  In the case  of the ultracold  Bose gas and using
the  Hartree-Fock-Bogoliubov formalism, the  $\lambda$-potential  permits to
self-consistently take into account the back action of the excitations
on the condensate in three-dimensional systems \cite{Vlambda} and also
in quasi-two-dimensional (2D) traps  \cite{HFB2D}. Scattering properties 
in low dimensional configurations are another powerful application of the
zero-range  potential approach.  For  example,  in the case of $s$-wave
scattering,  the  Bethe-Peierls  boundary condition   on  the  3D wave
function is a very efficient  way for finding the scattering amplitude
of two particles trapped in planar or linear atomic waveguides
\cite{QGLDPetrov,QGLDMaxim,Peano}. The important concept of tunable 
interaction in quasi-one-dimensional and quasi-two-dimensional systems
has      been  discovered     thanks       to      this       approach
\cite{QGLDPetrov,QGLDMaxim}. The few-body problem  is another area  of
research where the zero-range potential approach has already proven to
be very fruitful.  Exact results have  been  found for three and  four
particles  interacting  by      pairs   in  the   $s$-wave     channel
\cite{Petrov3B,Petrov3F,Petrovscatt,Mora}. These  studies are of first
importance in the context  of the resonant $s$-wave  scattering regime
obtained with  Feshbach resonances \cite{Ketterle_Feshbach}.  In these
experiments the scattering length can be tuned at  will, so that it is
possible to achieve the  BCS-BEC crossover of the two-component  Fermi
gas  \cite{Jochim,Zwierlein,Bourdel,Kinast,Greiner}.   The  zero-range
approach  permits us to evaluate the lifetime  (which is remarkably large
in the  neighborhood of the resonance) for  the weakly bound molecules
that appear in  the BEC  phase  and also to find    the value of   the
molecule-molecule  scattering  length    in terms   of  the   two-body
scattering length   \cite{Petrovscatt}.  For the unitary   quantum gas
(infinite two-body scattering  length) exact  results have been  found
for the few- and many-body problems \cite{Yvanscaling,Felix_OH}.  These
studies  show clearly  that the  zero-range potential   approach is an
efficient  tool beyond the    two-body physics and permits us to   obtain
non trivial properties in the few- and many-body problems.

More generally,  study of strongly  correlated systems obtained in the
vicinity of   scattering  resonances  is  actually  one  of  the  most
challenging directions in the field of ultracold atoms. These systems
are  not  only  interesting in  themselves  but  also  permit accurate
studies  of  issues  raised  in the  quantum many-body   problem.  The
BCS-BEC crossover in two-component Fermi gases is  an example of such
a situation  \cite{unitary,NSR,Randeria}.  The domain now  enlarges to
resonant scattering in $p$-wave
\cite{Regal,Ticknor,Zhang,Schunck,Chevy} and $d$-wave channels
\cite{Volz_1} allowing possible studies of the BCS-BEC crossover 
in channels of   high  angular momentum in   the near  future.   Major
interest in these  systems  stems from  the fact that  the strength of
correlations can  be tuned arbitrarily  while  the mean interparticle
distance remains large with respect to the range of interatomic forces
$R$ (for example, the $s$-wave  scattering length  can be adjusted  to
several  orders of magnitude larger   than $R$).  Consequently, one can
expect  that short range physics is  not directly  involved in a large
class  of many-body  properties which can  then be  described only  in
terms of the  low energy two-body behavior.  It  is worth pointing out
that  this fundamental   feature  has been  verified   in  the BCS-BEC
crossover  of the two-component   Fermi  gas.  For  this  reason,  the
zero-range  approach which is  parametrized  only by the two-body low
energy   physics is   a   very  appealing   tool for  studying   these
regimes. Actually, the state of the art  is as follows: broad $s$-wave
resonances (like the one in $^6$Li at 835 G \cite{Bourdel}) can be
accurately parametrized using  the Fermi pseudopotential, and  narrow
$s$-wave  resonance    \cite{Stenger}   can    be  studied     using a
generalization   of     the     Bethe-Peierls    boundary    condition
\cite{Petrov3B}. Recently, a general zero-range potential approach has
been introduced for  $p$-wave resonances \cite{pwave}  and  the aim of
this  paper  is  to extend  this   formalism  for a description of 
resonant scattering regime in arbitrary partial waves.
\subsection{Which generalization of actual pseudopotentials is needed?}
The idea of replacing a \emph{true} finite range two-body potential by
a zero-range   pseudopotential acting on  each  partial  wave has been
developed by Huang and Yang \cite{Huang} in the context
of  the hard-sphere   model. However, due to the importance of  the 
short range behavior in this  system, pseudopotentials describing  three-, 
four- and many-body correlations  were also proposed for modeling the 
system beyond the dilute  limit.   The actual situation  in ultracold
atoms  is radically different as  the  mean interparticle distance is
very much  larger than the range  of interatomic forces,  and two-body
pseudopotentials are sufficient   to describe the  low-energy behavior
for the many-body system even in strongly correlated regimes. Moreover
as the hard-sphere   model   cannot  support resonances,   the   tools
developed   in   Ref.~\cite{Huang}    cannot   be   used      for   our
purpose. Recently, a  generalization of the Fermi pseudopotential  for
arbitrary partial waves and scattering  phase shifts has been proposed
using   delta-shell  potentials in  the  limit   of small shell radius
\cite{Stock}, by       the    way  correcting      a    mistake     of
Ref.~\cite{Huang}.  Boundary conditions associated  with the zero-range
approach in Ref.~\cite{Stock} are defined for scattering states in free
space and depend explicitly  on their wave  number $k$.  Consequently,
this approach  applies only in time-independent situations.  Instead   in the present paper, we  show
that   in  resonant  regimes,  general  (that   is, energy independent)
boundary conditions  can     be found.    Another   feature  of    the
pseudopotentials  in Ref.~\cite{Stock} is  that their action is defined
on the radial  wave function of the  specified partial wave.  However,
it  is of   interest when dealing  with  wave  functions of  arbitrary
symmetry   (for example  in the   presence  of an  anisotropic   external
potential) to have the full expression of the pseudopotential, that is,
an  expression which contains  implicitly the projection operator over
the  interacting channel (denoted  in the following   by $\Pi_l$ for the
channel of angular quantum number~$l$).
\subsection{Organization of the paper}
The  paper is organized  as follows: in the first  part, the regime of
interest  (two-body  resonant    scattering)     is   introduced   and
characterized in each partial wave channel by a  set of two parameters
in the scattering phase shift. In the second part, the different tools
of the zero-range  approach in this regime are  derived.  We show that
the problem can be defined in terms of boundary conditions on the wave
function which generalize  the method introduced  by Bethe  and Peierls. 
We express these boundary conditions in spherical coordinates
and also in  Cartesian  coordinates  using  the symmetric trace   free
tensors used  in the usual  multipolar expansion. This way of defining
the zero-range approach  permits us to include   quite easily within  the
expression of the  pseudopotential both the projection  operator $\Pi_l$
and the $l$th derivatives of the  delta distribution. We show that in
each  interacting channel  there  exists a family  of pseudopotentials
generated by   a free parameter,  thus  generalizing the $\lambda$-potential
approach  \cite{Vlambda,pwave}.  The third  and last part, illustrates
the  formalism. We find   the  expression of the two-body   scattering
states from the  pseudopotential  and the Green's function  method. By
choosing a specific value  of the parameter  $\lambda$, this exact result is
also obtained  in the   first order  Born  approximation  for two-body
processes at finite energy. The pseudopotential of Ref.~\cite{Stock} is
recovered as a particular limit of  the present approach. Finally, the
scattering states and also the low-energy bound states are found to be
not mutually orthogonal in the zero-range scheme (except for the Fermi
pseudopotential). We introduce then a regularized scalar product which
solves  this inconsistency.  We  consider the  low-energy bound states
supported by the pseudopotential and show  that the new scalar product
leads to normalization constants which coincide with the results given
by the method based on the analyticy of the scattering amplitude
\cite{Landau}.
\section{Zero-range formalism in the resonant regime}
\subsection{Regime of interest}
Let us consider two particles in the  absence of an external potential
interacting via  a   short range  isotropic  potential  $U(\vec{r}\,)$
($\vec{r}\,$  being     the  relative coordinates    between   the two
particles).  We denote their reduced mass by $\mu$. In order to simplify
the  discussion and without loss of  generality, we work in the center-of-mass frame. The  asymptotic form  of  the scattering wave function
$\Psi_{\vec{k}}(\vec{r}\,)$ with   relative   wave vector $\vec{k}$   and
collisional   energy  $\displaystyle E   =\frac{\hbar^2k^2}{2\mu}$ defines the
partial  amplitudes  $f_l$ in  each  partial  wave channel  of angular
quantum number~$l$ through the relation:
\begin{equation}
\Psi_{\vec{k}}(\vec{r}\,) = \exp(i\vec{k}.\vec{r}\,)
+ \sum_{l=0}^\infty (2l\!+\!1) {\mathcal P}_l(\vec{n}.\vec{n}_k) f_l \frac{\exp(ikr)}{r} ,
\label{eq:asymptotic}
\end{equation}
where $kr \gg 1$, $\vec{n}=\vec{r}/r$, $\vec{n}_k=\vec{k}/k$ and ${\mathcal P}_l$
is  the Legendre polynomial  of degree $l$.  The partial amplitudes
can be also expressed  in   terms of the phase shift $\delta_l$ with:
\begin{equation}
f_l = \frac{1}{2ik} \left( \exp(2i \delta_l) - 1 \right) \quad . 
\label{eq:fl_deltal}
\end{equation}
Typical  situations   in  ultracold   atoms  corresponds  to   binary
collisions of  particles having small   enough relative velocities  to
ensure $kR \ll 1$,  where $R$ is the  potential range.  For  two neutral
alkali  atoms in    their   ground state,   the  asymptotic  form   of
$U(\vec{r}\,)$ is a  van der Waals  tail:  $U(\vec{r}\,)\simeq C_6/r^6$ and
the criterion for  low-energy scattering processes (collisional energy
$E \ll E_R$) is fixed by \cite{Dalibard,Marinescu}:
\begin{equation}
R = \left(\frac{2 \mu C_6}{\hbar^2}\right)^{1/4} \quad ; 
\quad E_R = \frac{\hbar^2}{2 \mu R^2} 
\end{equation}
Standard values for $R$  (and $E_R$) range from $\sim3.3~$nm 
for $^6$Li ($E_R \simeq 7.25$~mK) to $\sim 10~$nm for $^{133}$Cs 
($E_R \simeq  32.2~\mu$K). In this paper we consider situations where 
the phase shift can be parametrized in the following form:
\begin{equation}
k^{2l+1} \cot \delta_l(k) = - \frac{1}{w_l} - \alpha_l k^2 \quad .
\label{eq:phase_shift}
\end{equation}
Eq.(\ref{eq:phase_shift}) corresponds to the so-called effective  range
approximation  for a short range  potential \cite{Mott}: the parameter
$w_l$ characterizes the Wigner threshold regime  and $-2\alpha_l$ is ``the
effective range'' even if  it has the dimension  of a length only  for
$l=0$.     For neutral particles   interacting  with   a potential  of
asymptotic form $U(r) \simeq  C_n/r^n$ ($r \to \infty$) the threshold behavior  is
valid  for  $l<(n-3)/2$  and  the  effective range  approximation  for
$l<(n-5)/2$. In ultracold  atoms in their  ground state the power law
is  given by $n=6$  so that the expansion in Eq.(\ref{eq:phase_shift})
is  rigorous only in  the  $s$-wave channel  ($w_0$ is  the scattering
length), while  in the $p$-wave   channel only the  first  term in the
expansion is justified. For $l \geq 2$, even  the Wigner threshold regime
given in  Eq.(\ref{eq:phase_shift}) is not  valid.   However, we adopt
this form as a generic  way to parametrize the  resonant regime in an
arbitrary partial wave channel.

\begin{figure}
\resizebox{8cm}{!}{\includegraphics{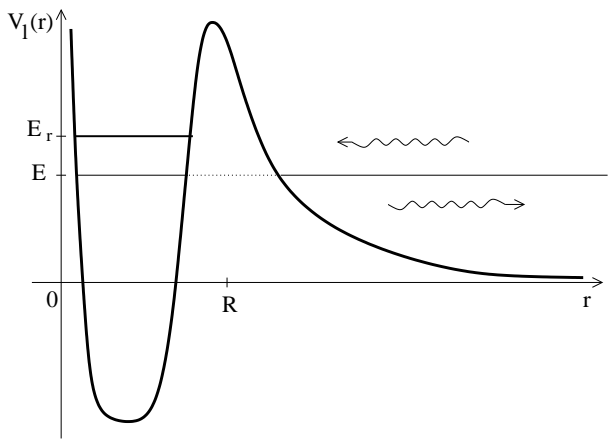}}
\caption{Schematic representation of a two-body effective radial potential 
$\displaystyle   V_l(r)=U(\vec{r})+\frac{\hbar^2\,  l(l+1)}{2\mu  r^2}$ in a
partial wave $l>0$.  $E_r$ is the energy of a quasi-bound state
\cite{Blatt,Landau} which   leads  to a  resonant behavior  in  the two-body
scattering for a collisional energy $E\simeq E_r$.}
\label{fig:quasi_BS}
\end{figure}

In broad $s$-wave resonances, the  effective range is negligible ($\alpha_0
\simeq 0$) and $|w_0|$ is very large as compared to the potential range $R$.
For $w_0>0$, $f_0$  has a pole associated  to a shallow bound state of
energy $-\hbar^2/2 \mu w_0^2$.  Its wave function  coincides (for $r>R$) with
the molecular state populated by pairs of particles  in the BEC region
of  the BEC-BCS  crossover. Remarkably,  in  the  neighborhood of  the
resonance, the scattering  cross section  takes large values  whatever
the    sign  of   the scattering length     $w_0$.    As explained  in
Ref.~\cite{Landau}, in the channels $l>0$,  this property is no  longer
verified, moreover the ``effective  range'' parameter is essential for
a modeling of the shape of the scattering amplitude and of the shallow
bound state  both. To define the  resonant regime for $l>0$, we follow
the  discussion given in  Ref.~\cite{Landau} and  use the expression of
the  partial      amplitude  associated with     the phase    shift in
Eq.(\ref{eq:phase_shift}):
\begin{equation}
f_l(k) = - \frac{ w_l k^{2l} }{1 + w_l \alpha_l k^2 + i w_l k^{2l+1}} \quad .
\label{eq:partial_amplitude}
\end{equation} 
The  resonant regime corresponds   to  situations where  the parameter
$|w_l|$ takes  arbitrary large values, so  that $|w_l \alpha_l| \gg R^2$.  It
can be   shown \cite{Landau} that  describing  the  resonant regime by
Eq.(\ref{eq:partial_amplitude})  is  valid   only for  $\alpha_l>0$   (this
assumption emerges also  as a consequence   of the regularized  scalar
product that we  introduce in the  last section).  The main difference
as compared to $s$-wave broad resonance is that the resonant character
of   the     scattering       cross section    associated         with
Eq.(\ref{eq:partial_amplitude})  depends crucially   on  the  sign  of
$w_l$.   This   property   can   be     easily  seen    by    writting
Eq.(\ref{eq:partial_amplitude}) in the Breit-Wigner form:
\begin{equation}
f_l = - \frac{1}{k} \frac{\Gamma/2}{ E-E_r + i \Gamma/2} \quad ,
\label{eq:Breit}
\end{equation}
with    $\displaystyle  E_r   =   -    \frac{\hbar^2}{2 \mu \alpha_l  w_l}$   and
$\displaystyle \Gamma  =  \frac{\hbar^2   k^{2l+1}}{\mu \alpha_l}$.     For large  and
positive values of $w_l$,  $E_r<0$ and the  scattering amplitude has a
pole  which  corresponds to  the presence  of  a real  bound  state of
vanishing  energy  $E_B \simeq E_r$.   In  this regime,  $|f_l|^2$ does not
present a resonant structure.   However, for large and  negative value
of  $w_l$,  the bound state transforms   into a long-lived quasi bound
state which produces a resonance in  the scattering cross section with
$|f_l|^2 \simeq 1/|w_l\alpha_l|$ in the neighborhood of $E = E_r$.  This feature
is a consequence  of the existence  of the centrifugal barrier  in the
radial  equation for  $l>0$ (see Fig.\ref{fig:quasi_BS}).  It is worth
pointing out that the resonant regime occurs for relative wave vectors
$k$  verifying $k R \ll 1$,   which  justifies  a zero-range   potential
approach.

In actual  experiments on ultracold  atoms, the resonant character in
the two-body     scattering is  ensured   using   Feshbach  resonances
\cite{Feshbach}.  Interaction   between  atoms  depends on    the spin
configuration of the two-body system and Feshbach resonances appear as
a result of the coupling between a closed  channel (supporting a short-range   
molecular state)  and  an  open  channel (associated  with the
scattering states).  The energy  of the short-range two-body molecular
state  can   be tuned by   use  of  an  external  magnetic  field. The
scattering resonance occurs when the  energy of the molecular state is
in the vicinity of the collisional energy of the two scattering atoms.
The molecular  state in  the closed channel depends on the short range
properties  of the  system  and is  not  described  by  the zero-range
approach.  Consequently, the   bound state or quasi-bound state  which
appears in the zero-range model is associated only with the tail of the
molecular state resulting from the  coupling between the closed and
open channels.

In  the  $s$-wave narrow    Feshbach  resonance  (large and   negative
effective range)  of  $^{23}$Na at $907$~G  \cite{Stenger}, $\alpha_0 \simeq
26$~nm, and $R \simeq 4.5$~nm.  In higher partial wave channels, scattering
is not isotropic \emph{i.e.} it depends  on the angular quantum number
$m$  of the  state.  For  example  the $p$-wave resonance in  $^{40}$K
\cite{Ticknor} ($R \simeq  6.8$~nm) has  been   parametrized with the  law
(\ref{eq:phase_shift})  in the states $|m|=0,1$  as a  function of the
external  magnetic   field.  There  are   two zero  energy  resonances
($E_r=0$) very close to one each other at $198.84$~G (for $m = 0$)
and  $198.31$~G  (for   $|m|=1$)    with a    width  characterized
respectively by $\alpha_1R \simeq 2.74$ and $2.81$.   Generalization of the zero
range approach for taking into  account this anisotropy adds technical
details  not essential  in a first  approach  and  in this  paper, the
discussion is centered on the modeling of isotropic interactions.
\subsection{Defining the zero-range approach}
In this section, the interacting potential $U(\vec{r}\,)$ which can be
neglected in the  Schr\"{o}dinger equation  for  $|\vec{r}\,| >  R$ is
replaced by a  specific behavior of the  wave function as the relative
coordinates between the   two interacting particles goes formally   to
zero.  This behavior defines boundary  conditions on the wave function
which depend on the parametrization of  the scattering phase shift in
Eq.(\ref{eq:phase_shift}).

Outside  the potential  range ($r>R$), the  scattering wave functions
$\{\Psi_{\vec{k}}(\vec{r}\,)\}$ of the two-body  problem are solutions of the
free Schr\"{o}dinger  equation  with the asymptotic behavior  given by
Eqs.(\ref{eq:asymptotic},\ref{eq:fl_deltal},\ref{eq:phase_shift}):
\begin{equation}
\Psi_{\vec{k}}(\vec{r}\,) = \sum_{l=0}^\infty  
{\mathcal P}_l(\vec{n}.\vec{n}_k) R_{kl}(r) \quad ,
\label{eq:psi_k}
\end{equation}
where $R_{kl}(r)$  are the radial functions which  can be expressed in
terms of the two spherical Bessel functions as~\cite{Abramowitz}:
\begin{equation}
R_{kl}(r) = {\mathcal N}_{kl} \left(\, j_l(kr) -  y_l(kr) \tan \delta_l(k)\, \right)   .
\label{eq:radial_free}
\end{equation}
The zero-range approach amounts to considering the formal continuation of
Eq.(\ref{eq:radial_free}) for ${0 \leq r \leq  R}$\,.  As a consequence, using
the behavior of $j_l(z)$ and $y_l(z)$ at the vicinity of the origin:
\begin{eqnarray}
&&j_l(z) = \frac{z^l}{(2l\!+\!1)!!}\left(1- \frac{z^2/2}{1!(2l\!+\!3)} + \dots \right)
\label{eq:serie_jz}\\
&&y_l(z) = -\frac{(2l\!-\!1)!!}{z^{l+1}}\left(1- \frac{z^2/2}{1!(1\!-\!2l)} + \dots \right)
\label{eq:serie_yz}
\quad ,
\end{eqnarray}
one  can  deduce the  scattering states  from   the following boundary
conditions of the wave function at $r=0$:
\begin{equation}
\lim_{r \to 0} \, \left(\frac{(2l\!-\!1)!!}{(2l)!!}\, \partial_r^{(2l+1)} 
- \frac{k^{2l+1}}{\tan \delta_l(k)} \right)\! r^{l+1} R_{kl}(r)   = 0 .
\label{eq:k_boundary}
\end{equation}
This expression  is a first  step toward  the formulation  of the zero
range approach, but  does not have a   general character in  the sense
that the quantum  number $k$ appears  explicitly.  Instead, we  search
for a boundary  condition which applies  on any linear combination  of
scattering states with different collisional  energies.  This way, the
formalism can be used when the collisional energy is not defined,
as it can happen in  time dependent situations. Using the parametrization
of the phase  shift  in  Eq.(\ref{eq:phase_shift}), one  obtains  from
Eq.(\ref{eq:k_boundary}):
\begin{equation}
\lim_{r \to 0} \, \left(\frac{(2l\!-\!1)!!}{(2l)!!}\, \partial_r^{(2l+1)} 
+ \alpha_l k^2 + \frac{1}{w_l} \right)\! r^{l+1} R_{kl}(r)   = 0 .
\label{eq:kk_boundary}
\end{equation}
The  idea   is then   to substitute   the  energy   dependent term  in
Eq.(\ref{eq:kk_boundary}) by the  appropriate radial derivative.  From
Eqs.(\ref{eq:serie_yz},\ref{eq:kk_boundary})  we find  the     desired
condition:
\begin{equation}
\lim_{r \to 0}({\mathcal D}_l+\frac{1}{w_l}) r^{l+1} R_{kl} = 0 \qquad ,
\label{eq:boundary}
\end{equation}
where we have introduced the differential operator:
\begin{equation}
{\mathcal D}_l =\frac{(2l\!-\!1)!!}{(2l)!!} \partial_r^{(2l+1)} 
+ (2l\!-\!1) \alpha_l \partial_r^2 \quad .
\label{eq:differential}
\end{equation}
We  are  now  ready to   define   the zero-range  approach in  general
situations.  The wave function $\Psi$ of  the interacting two-body system
is expanded in  each partial  wave channel in  terms  of the spherical
harmonics as:
\begin{equation}
\langle \vec{r}\,|\Pi_l|\Psi \rangle = \sum_{m=-l}^l 
\frac{c_\Psi^{\, lm}(r)}{r^{l+1}} Y^{lm}(\Omega) \quad ,
\label{eq:projected}
\end{equation}
where $\Pi_l$ is the projector on the subspace of angular quantum number
$l$, and $(r,\Omega)$ are the spherical  coordinates with $\Omega=(\theta,\phi)$. For $r
\neq 0$,  $\Psi(\vec{r}\,)$ is a solution of  the  free Schr\"{o}dinger equation
(it implies that the multipolar  radial functions $c_\Psi^{\, lm}(r)$ are
regular  functions of  the  interparticle distance   $r$),  and for an
isotropic interaction, the potential term is replaced by the following
boundary conditions on the wave function at $r=0$:
\begin{equation}
\lim_{r \to 0}\left({\mathcal D}_l+\frac{1}{w_l}\right) c_\Psi^{\, lm}(r) = 0 
\qquad 
-l \leq m \leq l \ .
\label{eq:boundary_general}
\end{equation}
Eq.(\ref{eq:boundary_general}) generalizes the  Bethe-Peierls approach
\cite{Bethe} which applies in the $s$-wave  channel when the effective
range can be neglected ($\alpha_0=0$):
\begin{equation}
\lim_{r \to 0} \frac{\partial_r (r \Psi)}{(r\Psi)} = -\frac{1}{w_0} \quad ,
\label{eq:Bethe_Peierls}
\end{equation}
where the  limit $|w_0| \gg R$  is associated  with  a zero energy broad
resonance.

Note that the generalization of Eqs.(\ref{eq:boundary_general}) to the
case of  an  anisotropic two-body  interaction is  straightforward  by
introducing   parameters that   depend   on  the  quantum  number~$m$:
$\{w_{lm},\alpha_{lm}\}$.

\subsection{Cartesian representation of the multipolar radial functions}

While the zero-range approach is  now established in  a general way by
Eqs.(\ref{eq:differential},\ref{eq:projected},\ref{eq:boundary_general}),
it  can  be  easier  to use    equivalent boundary  conditions  in the
Cartesian coordinate   system  (for   example   when the   system   is
inhomogeneous or anisotropic).  Furthermore,  as will be shown in  the
next section, using these conditions leads to a simple construction of
the  zero-range  pseudopotential (in particular,  derivatives  of  the
delta  distribution are  easily  introduced in Cartesian coordinates).
In the  following,   we  adopt the  standard   assumption  of implicit
summation   over  repeated indexes,  the  three   directions of  space
$\{x,y,z\}$  are labeled by a Greek  index $\alpha \in \{1,2,3\}$ and $[\alpha\beta\dots]$
denotes  a  set of $l$  indexes.   The idea   is to  use the symmetric
tensors ${\mathcal S}_l$ which appear in the multipolar expansion used
in electrostatics
\cite{Courant}. They can be defined through their components by:
\begin{equation}
{\mathcal S}_{l\,[\alpha\beta\dots]} = (-1)^l \frac{r^{l+1}}{(2l\!-\!1)!!} 
\partial\,^l_{[\alpha\beta\dots]} \frac{1}{r} \quad ,  
\label{eq:STF_Sl}
\end{equation}
where we have introduced  a short-hand notation for the $l$-th partial 
derivatives:
\begin{equation}
\partial\,^l_{[\alpha\beta\dots]}  =  \frac{\partial}{\partial x^\alpha} 
\frac{\partial}{\partial x^\beta} \dots  \quad ,
\label{eq:partial}
\end{equation}
and $\{x^\alpha\}$ are the three components of the radial vector $\vec{r}$ in
the Cartesian basis $\{ \vec{e}_\alpha \}$: \emph{i.e.}  $\vec{r} = x^\alpha
\vec{e}_\alpha$. These tensors can be expressed in terms of the component $n^\alpha$
of the unit vector $\vec{n}=\vec{r}/r$ and are functions of the angles
$\Omega$ only.  The  normalization factor in Eq.(\ref{eq:STF_Sl})  has been
chosen  for later convenience  and  explicit expressions of ${\mathcal
S}_l$   and   ${\mathcal  D}_l$     with  $l \leq   3$   are   given   in
Tab.(\ref{tab:Sl_Dl}).  In the  following, we   will use some   useful
properties of these tensors which  are  collected in the Appendix.  By
construction,  ${\mathcal  S}_l$   are traceless  ($\delta_{\alpha\beta}   {\mathcal
S}_l^{[\alpha\beta\dots]}  = 0$)  and symmetric  tensors of  rank $l$. They are
eigenvectors of   $\hat{L}^2$ ($\hat{L}$    is the angular    momentum
operator)   with   eigenvalue $\hbar^2 l     (l\!+\!1)$.  Similarly to the
spherical  harmonics $Y^{lm}(\Omega)$,   the tensors  ${\mathcal  S}_l$ can
serve as a  basis for  an alternative multipolar  expansion of  a wave
function, like in Eq.(\ref{eq:projected}). To this  end, we define the
Cartesian representation of the  multipolar radial functions  of order
$l$ associated  with a  wave function $\Psi$,  by a  symmetric trace free
tensor ${\mathcal C}_{\Psi,l}(r)$ having the components:
\begin{equation}
{\mathcal C}_{\Psi,l}^{[\alpha\beta\dots]}(r) = \frac{r^{l+1}}{l!}  \int_{\Omega}\! \frac{d^2\Omega}{4\pi} \, 
\left( {\mathcal S}_l^{[\alpha\beta \dots]} \Psi(r,\Omega)\right) \quad .
\label{eq:multipolar} 
\end{equation}
\begin{table}
\begin{ruledtabular}
\begin{tabular}{|c|c|c|}
\hline
$\ l \ $ & ${\mathcal S}_l^{[\alpha\beta\dots]}$ & ${\mathcal D}_l$ \\
\hline 
$0$ & $1$ & $\partial_r - \alpha_0 \partial_r^2$ \\
\hline
$1$ & $n^\alpha$ & $\frac{1}{2} \partial_r^3 + \alpha_1 \partial_r^2$ \\
\hline
$2$ & $n^\alpha n^\beta - \frac{1}{3} \delta^{\alpha\beta}$ & $\frac{3}{8} \partial_r^5 + 3 \alpha_2 \partial_r^2$ \\
\hline
$3$ & $n^\alpha n^\beta n^\gamma - \frac{1}{5} (n^\alpha \delta^{\beta\gamma}+ \mbox{permut.})$  
& $\frac{5}{16} \partial_r^7 + 5 \alpha_3 \partial_r^2$ \\
\hline
\end{tabular}
\end{ruledtabular}
\caption{Examples for the symmetric traceless 
tensors ${\mathcal S}_l$ Eq.(\ref{eq:STF_Sl}) and for the differential
operators  ${\mathcal D}_l$ in  Eq.(\ref{eq:boundary}). $n^\alpha$  are the
components of the unit vector $\vec{n}=\vec{r}/r$.}
\label{tab:Sl_Dl}
\end{table}
For example, the monopolar ($l=0$) function ${\mathcal C}_{\Psi,0}(r)$ is
just $r$  times the projection of  the  wave function in  the $s$-wave
channel.    Quite generally,   ${\mathcal C}_{\Psi,l}^{[\alpha\beta\dots]}(r)$ are
regular   functions  of $r$  analogous  to   the coefficients $c_\Psi^{\,
lm}(r)$  in Eq.(\ref{eq:projected})  and characterize  the behavior of
the wave  function $\Psi(r,\Omega)$ in the   channel~$l$ through the expansion
(see Eqs.(\ref{eq:Pl_Sl},\ref{eq:projector_Pl}) in the Appendix):
\begin{equation}
\langle \vec{r}\,| \Pi_l | \Psi \rangle = \frac{(2l\!+\!1)!!}{r^{l+1}} \, 
 {\mathcal S}_{l [\alpha\beta \dots]} {\mathcal C}_{\Psi,l}^{[\alpha\beta\dots]}(r) \quad .
\label{eq:projector}
\end{equation}
Using these notations,  the generalized Bethe-Peierls boundary conditions
Eq.(\ref{eq:boundary_general}) can be written as:
\begin{equation}
\lim_{r \to 0}({\mathcal D}_l+\frac{1}{w_l})\, {\mathcal C}_{\Psi,l}^{[\alpha\beta\dots]}(r) = 0 \quad .
\label{eq:boundary_tensorial}
\end{equation}
Eqs.(\ref{eq:multipolar},\ref{eq:projector},\ref{eq:boundary_tensorial})
are equivalent  to  Eqs.(\ref{eq:projected},\ref{eq:boundary_general}).
We show in the next section that they permit to find a general 
form of the pseudopotential.

Note  that     an  alternative   formulation     of   the    condition
(\ref{eq:boundary_tensorial})  in  the  $s$-wave   channel  amounts to
imposing the following singularity on the two-body wave function:
\begin{equation}
\langle\vec{r}\,|\Pi_0| \psi  \rangle =  \frac{{\mathcal C}_{\Psi,0}(0)}{r} \left(1-\frac{r}{a} + A_0 ( r^2 + 2 \alpha_0 r ) \right) + 
{\mathcal O}(r^3) .
\end{equation}
In the $p$-wave channel $l=1$, the short-range behavior of the interacting 
two-body wave function takes also a simple form:
\begin{equation}
\langle  \vec{r}\,|\Pi_1| \Psi \rangle = \frac{\vec{p}.\vec{r}}{r^3}\! 
\left( 1 - \frac{r^3}{3{\mathcal V}_s} + A_1 ( 3 r^2 - 2 \alpha_1 r^3) \right)  
\!+ {\mathcal O}(r^2) ,
\end{equation}
where the ``dipolar momentum'' $\vec{p}$ defined by $p^\alpha = 3 {\mathcal
C}_{\Psi,1}^\alpha(0)$, characterizes   the $p$-wave singularity  of  the wave
function.    In    general         situations     the     coefficients
($A_0$,$A_1$,${\mathcal C}_{\Psi,0}(0)$, $\vec{p}\,$) can be functions of
time and of the center of mass coordinates of the two particles.

\subsection{Zero-range pseudopotential}

Instead of giving a final expression  of the pseudopotential directly,
we prefer to detail the different steps in its derivation.  Such a way
of presentation can be useful if one  wants to derive pseudopotentials
associated with    others  expressions   of    the  phase  shift    in
Eq.(\ref{eq:phase_shift}) or in different contexts (low dimensions for
example). In either cases, a zero-range pseudopotential can be deduced
from the three following constraints: ($i$) it cancels the delta terms
arising from action of the Laplacian on  the singularities of the wave
function at $r=0$ in  the Schr\"{o}dinger equation, ($ii$)  it imposes
the correct boundary conditions and ($iii$) it can be used in the Born
approximation.   The first  condition (item-$i$)  can  be fulfilled by
introducing potential terms in the  different channels $l$ of the form
(see Eq.(\ref{eq:singularity}) in the Appendix):
\begin{eqnarray}
\langle \vec{r}\, | V_{l}^{(i)} | \Psi \rangle 
=&&\!\!\!\!\!\! (-1)^{l+1}\, \frac{2\pi(2l\!+\!1)\hbar^2}{\mu} \nonumber \\ 
&& {\mathcal C}_{\Psi,l}^{[\alpha\beta\dots]}(0)\, 
(\partial\,^l_{[\alpha\beta\dots]} \delta)(\vec{r}\,)
\label{eq:pseudo(i)} \quad .
\end{eqnarray}
In  order        to      include   the       boundary       conditions
Eqs.(\ref{eq:boundary_tensorial})   (item-$ii$), we   consider     the
``boundary operator'' ${\mathcal B}_l$ defined by:
\begin{eqnarray}
\langle \vec{r}\, | {\mathcal B}_l | \Psi \rangle = 
(\partial\,^l_{[\alpha\beta\dots]} \delta)(\vec{r}\,) 
\lim_{r \to 0}({\mathcal D}_l+\frac{1}{w_l})\,{\mathcal C}_{\Psi,l}^{[\alpha\beta\dots]}(r)
\label{eq:pseudo(ii)} .
\end{eqnarray}
By construction wave functions  belonging to  the kernel of  ${\mathcal
B}_l$ satisfy the desired   boundary  conditions. Consequently, 
the  family   of pseudopotentials defined by:
\begin{equation}
V_l^{(i)} + \beta\, {\mathcal B}_l \quad ,
\label{eq:beta_potential}
\end{equation}
where $\beta$ is an arbitrary parameter, can be used to solve the two-body
problem in the resonant regime.  Interestingly, while exact zero-range
solutions can be obtained  $\forall \beta \neq 0$, it  is not hard to be  convinced
that a  first order  Born approximation  performed for example  in the
computation of the two-body scattering states leads to a $\beta$-dependent
result.   Indeed in  this  approximation, the  two-body  wave function
$\Psi^{(0)}$  is an eigenstate   of the   non-interacting Schr\"{o}dinger
equation (a plane wave  in the example  chosen) and is then a  regular
wave   function of   the  radius  $r$.  Consequently,  the   action of
$V_l^{(i)}$   on $\Psi^{(0)}$ gives   a zero  result while  $\langle\vec{r}\, |
V_l^{(i)}   + \beta\,  {\mathcal  B}_l   |\Psi^{(0)} \rangle$  is    $\beta$-dependent.
Furthermore, and  we  will show this  property explicitly  in the next
section, varying $\beta$ offers the freedom  to adjust the pseudopotential
in  such a way  that the first order Born  approximation is exact at a
given collisional energy  (item-$iii$). Finally, the expression of the
pseudopotential  in Eq.(\ref{eq:beta_potential}) although general, can
be rewritten in the more convenient following form:
\begin{equation}
\langle \vec{r}\, | V_{\lambda,l} | \Psi \rangle = (-1)^l  g_{\lambda,l} 
(\partial\,^l_{[\alpha\beta\dots]} \delta)(\vec{r}\,) 
{\mathcal R}_{\lambda,l}^{[\alpha\beta \dots]}[\Psi]
\label{eq:pseudopotential} \quad .
\end{equation}
In  this  expression,  $g_{\lambda,l}=(-1)^l \beta$ is the coupling constant 
defined by:
\begin{equation}
\displaystyle g_{\lambda,l} = \frac{2 \pi (2l\!+\!1) \hbar^2 w_l }{\mu (1-\lambda w_l)} \quad ,
\label{eq:g_lambda}
\end{equation}
and ${\mathcal R}_{\lambda,l}[\,.\,]$ are symmetric trace free tensors which
generalize    the      regularizing   operator     introduced       in
Refs.\cite{Vlambda,pwave}. They act on a wave function $\Psi$ as:
\begin{equation}
{\mathcal R}_{\lambda,l}^{[\alpha\beta \dots]}[\Psi] = \lim_{r \to 0}  
({\mathcal D}_{l} + \lambda ) {\mathcal C}_{\Psi,l}^{[\alpha\beta \dots]}(r) 
\label{eq:regularizing} \quad .
\end{equation}
The    parameter   $\lambda$     appears   in Eq.(\ref{eq:g_lambda})     and
Eq.(\ref{eq:regularizing})  both,  and by   construction (like  $\beta$ in
Eq.(\ref{eq:beta_potential})) it is  a  free parameter of the  theory:
exact  wave  functions  obtained  in  the zero-range formalism  do not
depend     on  the   value     of~$\lambda$.    The   pseudopotentials    in
Eq.(\ref{eq:pseudopotential}) generalize  for arbitrary   partial wave
channels the $s$-wave  and $p$-wave  $\lambda$-potentials already introduced
in Ref.~\cite{Vlambda,pwave}.

\section{Some illustrations and properties of the pseudopotential}

\subsection{Two-body scattering states from the Green's method}

This section  is both an  illustration of the  formalism and a  way to
test   the   proposed  form    of   the    pseudopotential   given  in
Eq.(\ref{eq:pseudopotential}).    We consider   the   case   where two
particles interact in the $l$th partial wave channel without external
potential and we derive the scattering states from the general Green's
function   method.      The  interacting  two-body     wave   function
$\Psi_{\vec{k}}(\vec{r}\,)$ (wave vector $\vec{k}$, energy $E=\hbar^2k^2/2\mu$)
is then the solution of the integral equation:
\begin{equation}
\Psi_{\vec{k}}(\vec{r}\,) = \Psi^{(0)}_{\vec{k}}(\vec{r}\,) - \int\!\!d^3\vec{r}\,'\, 
G_E(\vec{r},\vec{r}\,') 
\langle \vec{r}\,'\, |V_{\lambda,l}| \Psi_{\vec{k}} \rangle \quad ,
\label{eq:green}
\end{equation}
where $\Psi^{(0)}_{\vec{k}}(\vec{r}\,) = \exp(i\vec{k}.\vec{r}\,)$  is the 
incident plane wave and $G_E$  is the one-body outgoing free Green's function 
at energy $E$:
\begin{equation}
G_E(\vec{r},\vec{r}\,') = \frac{\mu}{2 \pi \hbar^2} 
\frac{\exp i ku}{u} \quad \mbox{with} \quad u = | \vec{r}-\vec{r}\,'| .
\label{eq:green_function}
\end{equation}
Integrating Eq.(\ref{eq:green}) over $\vec{r}\,'$  coordinates gives:
\begin{eqnarray}
\Psi_{\vec{k}}(\vec{r}\,) = \Psi^{(0)}_{\vec{k}}(\vec{r}\,) - 
&&\!\!\!\!\!\! \frac{\mu g_{\lambda,l}}{2\pi \hbar^2} 
{\mathcal R}_{\lambda,l}^{[\alpha\beta \dots]}[\Psi_{\vec{k}}]  \nonumber \\
&& \quad  \left(\partial\,'^l_{[\alpha\beta\dots]} \frac{\exp i ku}{u}\right)_{r'=0} ,
\label{eq:integration1}
\end{eqnarray}
where $\partial\,'^l_{[\alpha\beta\dots]}$ stands for  the $l$th  partial derivatives
with respect  to $\vec{r}\,'$  coordinates.  Thanks to the  trace free
property  (${\delta_{\alpha\beta}{\mathcal R}_{\lambda,l}^{[\alpha\beta  \dots]}[\Psi_{\vec{k}}]=0}$),
contribution  of the  $l$th derivative in  Eq.(\ref{eq:integration1})
takes a simple form. Indeed, for a given function $F(u)$, one can show
that:
\begin{eqnarray}
&&\!\!\!\!\!\! \left(\partial\,'^l_{[\alpha\beta\dots]} F(u) \right)_{r'=0} 
{\mathcal R}_{\lambda,l}^{[\alpha\beta \dots]}[\Psi_{\vec{k}}]=\nonumber\\ 
&& \quad  (n_\alpha n_\beta \dots) 
{\mathcal R}_{\lambda,l}^{[\alpha\beta \dots]}[\Psi_{\vec{k}}] 
(-r)^l \left(\frac{1}{r}\partial_r\right)^l F(r)
 \ ,
\end{eqnarray}
and with $F(u)=G_E(\vec{r},\vec{r}\,')$, Eq.(\ref{eq:integration1}) reads 
finally: 
\begin{eqnarray}
\Psi_{\vec{k}}(\vec{r}\,) &=&  \Psi^{(0)}_{\vec{k}}(\vec{r}\,) 
- \frac{\mu g_{\lambda,l}}{2\pi \hbar^2} 
{\mathcal S}_{l [\alpha\beta \dots]} 
{\mathcal R}_{\lambda,l}^{[\alpha\beta \dots]}[\Psi_{\vec{k}}] \nonumber \\
&&\qquad (-r)^l \left(\frac{1}{r}\partial_r\right)^l\!\left(\frac{\exp(ikr)}{r}\right) .
\label{eq:green_zerorange}
\end{eqnarray}
As $r\to 0$, one recognizes a  multipolar singularity in the two-body wave function:
\begin{equation}
\Psi_{\vec{k}}(\vec{r}\,) = - \frac{(2l\!+\!1)!!\,w_l}{(1-\lambda w_l)} 
\frac{{\mathcal S}_{l [\alpha\beta \dots]}}{r^{l+1}} 
{\mathcal R}_{\lambda,l}^{[\alpha\beta \dots]}[\Psi_{\vec{k}}] 
+ {\mathcal O}(r^{1-l}) ,
\end{equation}
and comparison with Eq.(\ref{eq:projector}) gives the multipole components:
\begin{equation}
{\mathcal C}_{\Psi_{\vec{k}},l}^{[\alpha\beta\dots]}(0) = -\frac{w_l}{1-\lambda w_l} 
{\mathcal R}_{\lambda,l}^{[\alpha\beta \dots]}[\Psi_{\vec{k}}] \quad .
\end{equation}
In the partial wave channels of  angular quantum number different from
$l$, components of the wave function $\Psi_{\vec{k}}(\vec{r}\,)$ coincide
as   expected        with        the       partial     waves        of
$\Psi_{\vec{k}}^{(0)}(\vec{r}\,)$. All the information on the interaction
are gathered  in  the term  ${{\mathcal  S}_{l  [\alpha\beta \dots]}  {\mathcal
R}_{\lambda,l}^{[\alpha\beta\dots]}[\Psi_{\vec{k}}]}$  in Eq.(\ref{eq:green_zerorange}).
From Eq.(\ref{eq:projector}), one can see  that it contains implicitly
the projector $\Pi_l$ through:
\begin{eqnarray}
&&\!\!\!\!\!\! {\mathcal S}_{l [\alpha\beta \dots]}{\mathcal R}_{\lambda,l}^{[\alpha\beta \dots]}[\Psi] 
 =   \nonumber\\
&&\quad  \frac{1}{(2l\!+\!1)!!}\,\lim_{r \to 0}  ({\mathcal D}_{l} + \lambda ) r^{l+1} 
\langle \vec{r}\,| \Pi_l | \Psi \rangle  \ .
\label{eq:reg_proj_psi}
\end{eqnarray}
A closed  equation  is obtained  by applying the  operator ${{\mathcal
S}_{l [\alpha\beta \dots]} \mathcal R}_{\lambda,l}^{[\alpha\beta \dots]}[\,.\,]$ on both sides
of Eq.(\ref{eq:green_zerorange}).  Using the  usual decomposition of a
plane wave over spherical ones:
\begin{eqnarray}
\exp(i\vec{k}.\vec{r}\,) = \sum_{l=0}^\infty i^l (2l\!+\!1) {\mathcal P}_l(\vec{n}.\vec{n}_k) j_l(kr) \ ,
\label{eq:kzjr} 
\end{eqnarray}
together with Eq.(\ref{eq:serie_jz}) this leads to:
\begin{eqnarray}
\lim_{r \to 0}  ({\mathcal D}_{l} + \lambda ) r^{l+1} 
&&\!\!\!\!\!\! \langle \vec{r}\,| \Pi_l | \Psi^{(0)}_{\vec{k}} \rangle  = \nonumber\\
&&(2l\!+\!1)!!\,(ik)^l {\mathcal P}_l(\vec{n}.\vec{n}_k) \quad,
\end{eqnarray}
on the other hand:
\begin{eqnarray}
\lim_{r \to 0}  ({\mathcal D}_{l} + \lambda )&&\!\!\!\!\!\! r^{2l+1} \left(\frac{1}{r}\partial_r\right)^l\!
\left(\frac{\exp(ikr)}{r}\right)  = \nonumber \\ 
&&\!\!\!\!\!\!(-1)^l (2l\!-\!1)!!\, (\lambda +\alpha_l k^2 + i k^{2l+1}) \ ,
\end{eqnarray}
and finally, one gets:
\begin{eqnarray}
 {\mathcal S}_{l [\alpha\beta \dots]}{\mathcal R}_{\lambda,l}^{[\alpha\beta \dots]}[\Psi_{\vec{k}}]=
\frac{(1-\lambda w_l)(ik)^l {\mathcal P}_l(\vec{n}.\vec{n}_k)}{1+w_l \alpha_l k^2 +i w_l k^{2l+1}}\ .
\end{eqnarray}
As expected,  the expression of  the scattering wave function does not
depend on the parameter $\lambda$:
\begin{eqnarray}
\Psi_{\vec{k}}(\vec{r}\,) &=&  \Psi^{(0)}_{\vec{k}}(\vec{r}\,) -  
\frac{(2l\!+\!1) w_l {\mathcal P}_l(\vec{n}.\vec{n}_k)}{1+w_l \alpha_l k^2 +i w_l k^{2l+1}}
\nonumber \\
&&\qquad (-ikr)^l \left(\frac{1}{r}\partial_r\right)^l\!\left(\frac{\exp(ikr)}{r}\right)\ .
\label{eq:scattering_state}
\end{eqnarray}
Asymptotically for $kr \gg 1$, the two-body wave function is given by:
\begin{equation}
\Psi_{\vec{k}}(\vec{r}\,) = \exp(i\vec{k}.\vec{r}\,) + (2l\!+\!1) 
{\mathcal P}_l(\vec{n}.\vec{n}_k) f_l \frac{\exp(ikr)}{r},
\end{equation}
where  the partial  amplitude $f_l$ is given by Eq.(\ref{eq:partial_amplitude}). 

It is  interesting to note that the  exact result can be also obtained
at the level   of the  first order   Born approximation  (which is   a
``$\lambda$-dependent'' treatment) by   choosing a particular value of   the
parameter $\lambda$:
\begin{equation}
\lambda^{Born} = - \alpha_l k^2 - i k^{2l+1} \quad .
\label{eq:lambda_Born}
\end{equation}
For example by setting $\lambda=0$ in a first order Born approximation gives
a correct   evaluation of a scattering   process at a vanishly small
collisional energy (Wigner threshold).

Generalization to the case where the interaction occurs in several partial
wave channels is straightforward: it is obtained after simple addition 
of the corresponding pseudopotentials given by Eq.(\ref{eq:pseudopotential}).

\subsection{Links with other formalisms}

It  is instructive to compute  the matrix element of $V_{\lambda,l}$ between
two states  $|\Phi\rangle$ and $|\Psi \rangle$.  A  simplifying step is  to realize that
the projection operator   in the  partial   wave ($l$) is   implicitly
present      in    the     expression  of     the      pseudopotential
Eq.(\ref{eq:pseudopotential}). Indeed, assuming that $\Phi(\vec{r}\,)$ is
regular at $r=0$, one obtains:
\begin{equation}
 {\mathcal S}_l^{[\alpha\beta\dots]}(\partial\,^l_{[\alpha\beta\dots]} \Phi)_{r=0} 
= l! \, \lim_{r\to 0} \frac{\langle \vec{r}\,|\Pi_l|\Phi\rangle}{r^l} \quad .
\label{eq:proj_phi}
\end{equation}
Moreover from  Eqs.(\ref{eq:reg_proj_psi},\ref{eq:proj_phi}) and
Eq.(\ref{eq:thorne}) in the Appendix, one finds:
\begin{eqnarray}
\langle \Phi | V_{\lambda,l} |\Psi \rangle = &&\!\!\!\!\!\!  g_{\lambda,l}  \int \!\! \frac{d^2\Omega}{4\pi}
\, \biggl( \lim_{r \to 0} \frac{\langle \Phi | \Pi_l |\vec{r} \, \rangle}{r^l} 
\nonumber\\
&& \lim_{r \to 0} ({\mathcal D}_l + \lambda ) r^{l+1}\langle \vec{r}\, | \Pi_l |\Psi \rangle \biggl)
\quad . 
\label{eq:matrix_elements}
\end{eqnarray}
\indent First, we consider plane waves ${\langle \vec{r} \,| \vec{k} \,\rangle =
\exp(i\vec{k}.\vec{r})}$. The matrix element between two such states is:
\begin{equation}
\langle \vec{k}\,' |  V_{\lambda,l} |\vec{k}\,\rangle =  g_{\lambda,l} (k k\,')^l \, {\mathcal
P}_l(\vec{n}_k\,'.\vec{n}_k) \quad ,
\label{eq:Born_tmatrix}
\end{equation}
with $\vec{n}_k\,'=\vec{k}'/k'$. For     wave    vectors     of    same     modulus     ($\displaystyle
k=k\,'=\sqrt{\displaystyle  2\mu E}/\hbar$) and for  the particular value of
the   free  parameter  $\lambda$     given  in Eq.   (\ref{eq:lambda_Born}),
comparison  with   the expression  of    the  partial wave   amplitude
Eq.(\ref{eq:partial_amplitude}) shows  that Eq.(\ref{eq:Born_tmatrix})
coincides  with the  on-shell  $T$-matrix  $t(\vec{k}\,' \gets \vec{k}\,)$
taken at energy $E$:
\begin{equation}
t(\vec{k}\,'\! \gets \vec{k}\,) = \langle \vec{k}\,' | V_{\lambda^{Born},l} |\vec{k} \,\rangle 
= -\frac{2\pi \hbar^2}{\mu} f(\vec{k}\,'\! \gets \vec{k}\,) , 
\end{equation}
where $f(\vec{k}\,'\! \gets  \vec{k}\,) = (2l\!+\!1)  {\mathcal P}_l(\vec{n}_k\,'.\vec{n}_k)  f_l(k)$  is  the full scattering
amplitude.

Another interesting application of Eq.(\ref{eq:matrix_elements}) is 
given by considering spherical outgoing waves:
\begin{equation}
\langle \vec{r} \,| k\,l\,m\rangle = Y^{lm}(\Omega) j_l(kr) \quad .
\end{equation}
In this case, the matrix elements are given by:
\begin{equation}
\langle k\,l\,m | V_{\lambda,l} | k\,'\,l\,'\,m\,' \rangle = 
\delta_{ll'} \delta_{m m\,'} \frac{g_{\lambda,l}  (k k\,')^l }{4\pi (2l\!+\!1)} 
\quad .
\label{eq:lambda_roth}
\end{equation}
For  $\lambda=0$,  this     result  coincides with    the   one  obtained in
Ref.~\cite{Roth}.  Here, the $\lambda$-freedom  is a new ingredient and shows
clearly that mean field calculations in  zero-range approaches have to
be  developed carefully.   For   example, considering a  two-component
Fermi gas close to the unitary  limit with  $\alpha_0=0$ and a large and 
negative value of the scattering length $w_0$, the mean
field model  in Ref.~\cite{Roth} predicts an   instability which is not
observed in experiments \cite{Regal,Ticknor,Zhang,Schunck,Chevy}. This
failure  can be explained  as follows. The  interaction concerns only
particles  close to the  Fermi surface (Pauli   blocking) with a Fermi
wave vector $k_F$. However, in this mean field approach the approximation
made on the interaction term is only valid for binary processes of
vanishly   small  colliding energies (equivalent to a Born approximation
with the pseudopotential and the particular choice $\lambda=0$). 
Consequently in  the resonant regime  where $k_F |w_0| \gtrsim 1$, 
the characteristic collisional energy of pairs of particles on the 
Fermi surface is $\hbar^2/m w_0^2$ and the approximation made is not 
justified. By the way, it is possible to build  a qualitative and  
simple mean field  model which supports the stability of the gas at 
the BCS-BEC crossover~\cite{Box}.

Taking   two states with   angular  parts  characterized  by the  same
spherical harmonic  in Eq.(\ref{eq:matrix_elements}), one  obtains the
expression of the pseudopotential in radial coordinates:
\begin{equation}
v_{\lambda,l} = \frac{g_{\lambda,l}}{4\pi} \frac{\delta(r)}{r^{l+2}} 
\lim_{r \to 0} ({\mathcal D}_l+\lambda) (r^{l+1} \, . \, )
\label{eq:pseudo_radial} \quad .
\end{equation}
In the particular case $\lambda=0$, $\alpha_l=0$, {\emph r.h.s.} of 
Eq.(\ref{eq:pseudo_radial}) coincides with the form of the 
pseudopotential introduced in Ref.~\cite{Stock}:
\begin{equation}
v_{0,l} = \frac{\hbar^2 w_l (2l\!+\!1)!!}{2 \mu (2l)!!} 
\frac{\delta(r)}{r^{l+2}} \lim_{r \to 0} \partial_r^{2l+1} ( r^{l+1}\, . \, ) \quad .
\label{eq:pseudo_stock}
\end{equation}
Note   that the delta-shell   regularization  procedure  introduced in
Ref.~\cite{Stock}  while correct appears not   to  be essential in  the
zero-range approach.

\subsection{Regularized scalar product}

Except for the case of the Fermi pseudopotential which is a particular
case  of   Eq.(\ref{eq:pseudopotential}),   two   scattering    states
$\Psi_{\vec{k}\,'}$ and  $\Psi_{\vec{k}}$ (with ${\vec{k} \neq \vec{k}\,'}$) of
expressions given by Eq.(\ref{eq:scattering_state}) are not orthogonal
each others with respect to the usual  scalar product. To examine this
point,   we exclude the singularity   at $r=0$ in   the scalar product
between the   two   different  scattering states    by   introducing a
cut-off~$r_0$. We obtain from the Schr\"{o}dinger equation:
\begin{eqnarray}
&&\!\!\!\!\!\! \int_{r>r_0}\!\!\!\!\!\!\!\! d^3\vec{r} \  
\Psi_{\vec{k}\,'}^*(\vec{r}\,) \Psi_{\vec{k}}(\vec{r}\,) 
= \frac{r_0^2}{k^2-k\,'^2} \nonumber\\
&&\int_{r=r_0}\!\!\!\!\!\!\!\! d^2\Omega \, 
\Psi_{\vec{k}\,'}^*(\vec{r}\,) \partial_r \Psi_{\vec{k}}(\vec{r}\,) -  
\Psi_{\vec{k}}(\vec{r}\,) \partial_r \Psi_{\vec{k}\,'}^*(\vec{r}\,) \quad .
\label{eq:usual_scalar_product}
\end{eqnarray}
This  relation is general \emph{i.e.} it  is also valid for scattering
wave functions  associated   with a finite  range   two-body potential
whatever the  value  of  $r_0$.   In  standard  situations where   the
Hamiltonian   is     hermitian,        the     {\it   r.h.s.}       of
(\ref{eq:usual_scalar_product}) is  zero for  $r_0=0$.  However in the
zero-range approach, for   wave   functions satisfying  the   boundary
conditions in Eq.(\ref{eq:boundary}),   as   ${r_0 \to  0}$  this   term
converges toward a constant value for ${l=0}$ and ${\alpha_0 \neq 0}$ and even
worse, it  diverges for ${l>0}$.  Interestingly,  the  regular part of
Eq.(\ref{eq:usual_scalar_product})  for  such wave  functions  can  be
obtained through the following surface integration:
\begin{eqnarray}
&&\!\!\!\!\!\!\!\! \operatornamewithlimits{Reg}_{r_0 \to 0} \left\{ \int_{r>r_0}\!\!\!\!\!\!\!\! 
d^3\vec{r} \ \Psi_{\vec{k}\,'}^*(\vec{r}\,) 
\Psi_{\vec{k}}(\vec{r}\,) \right\} =  \nonumber\\
&&\!\!- \lim_{r_0 \to 0} \sum_{l=0}^\infty 
\frac{\alpha_l\, r_0^{2l+2}}{((2l\!-\!1)!!)^2} \!\!\! \int_{r=r_0}\!\!\!\!\!\!\!\!\!\!
d^2\Omega \,  \Psi_{\vec{k}\,'}^*(\vec{r}\,) \langle \vec{r}\,|\Pi_l|\Psi_{\vec{k}}\rangle  ,
\label{eq:reste}
\end{eqnarray}
where  the operator $\displaystyle \operatornamewithlimits{Reg}_{r_0 \to
0}\left\{\,.\,\right\}$  extracts the   regular part at  ${r_0=0}$.  The
meaning of this result is that the Laplacian operator is not Hermitian
with respect  to the usual scalar product  when one considers singular
functions like the  ones in Eq.(\ref{eq:scattering_state}). In another
hand,   the \emph{true} scattering  states  associated with the finite
range  potential experienced by the  particles, are orthogonal to each
other so that  direct  identification  of the  \emph{true}  scattering
states   with  the   states    $\left\{|\Psi_{\vec{k}}\rangle\right\}$  is    not
possible. This feature follows from  the fact that the mapping between
the two eigen  basis is not valid  for  $r \lesssim R$ while  it is justified
only  outside  the  potential  range.  Indeed, the  singular  boundary
behavior imposed on  wave  functions in Eq.(\ref{eq:boundary_general})
is a   way to reproduce the effect   of the  \emph{true}  finite range
potential  for ${r>R}$ but has   a formal character for  interparticle
distances $r \lesssim R$. In order to  understand this issue more closely, we
consider the   sphere of radius $R$   which  delimits the  outer parts
(region $r>R$) from the inner parts (region  $r<R$) of the \emph{true}
wave  functions.  Contribution to the  scalar  product coming from the
outer-part   of   the  \emph{true}   wave    functions  is   given  by
Eq.(\ref{eq:usual_scalar_product})      with  $r_0=R$    and    due to
orthogonality   between  the  \emph{true}  scattering  states,  scalar
product  of  the   inner-parts cancels   this  term.  However  in  the
zero-range potential approach the inner  parts  of the wave  functions
($r<R$) are  not properly described,  so that compensation of  the two
contributions does not occur. To solve this inconsistency, the idea is
to modify the usual scalar product  itself by including implicitly the
contribution of the inner  parts. We define  then a regularized scalar
product   $(\,.\,|\,.\,)_0$   by    subtracting   the   surface   term
(\emph{r.h.s.}  of  Eq.(\ref{eq:usual_scalar_product})) from the usual
scalar product as ${r_0 \to 0}$:
\begin{eqnarray}
(\Psi &&\!\!\!\!\!\!  | \Phi)_{0} = 
\operatornamewithlimits{Reg}_{r_0 \to 0} 
\left\{ \int_{r>r_0}\!\!\!\!\!\! d^3\vec{r} \,  
\Psi^*(\vec{r}\,) \Phi(\vec{r}\,) \right\} \nonumber \\
&&\!\!\!\!\!\!\!\!  + \lim_{r_0 \to 0} \sum_{l=0}^\infty \frac{\alpha_l\, 
r_0^{2l+2}}{((2l\!-\!1)!!)^2} \!\!\! 
\int_{r=r_0}\!\!\!\!\!\! d^2\Omega \,  
\Psi^*(\vec{r}\,) \langle \vec{r}\,|\Pi_l | \Phi \rangle  ,
\label{eq:regu_sp}
\end{eqnarray}
so that by construction, two eigenstates of the zero-range Hamiltonian
are orthogonal to each other with respect to this metrics.

In   general there    are    several  possible   representations    of
Eq.(\ref{eq:regu_sp})  due   to  the various  and  equivalent  ways of
extracting  a regular part. We  give below explicit expressions of the
regularized  scalar  product   associated  with  the   pseudopotential
(\ref{eq:pseudopotential})  in  the    $l=0$ and  $l=1$ partial   wave
channels:
\begin{itemize}
\item  Particles interacting in the $s$-wave channel only:
\begin{eqnarray}
(\Psi|\Phi)_{0} = &&\!\!\!\!\!\! \lim_{r_0 \to 0} \left\{
\int_{r>r_0}\!\!\!\!\!\! d^3\vec{r} \,  \Psi^*(\vec{r}\,) \Phi(\vec{r}\,)  
\right. \nonumber\\
&&
\left.  + \alpha_0 r_0^2  \int_{r=r_0}\!\!\!\!\!\! d^2\Omega \,  \Psi^*(\vec{r}\,) 
\Phi(\vec{r}\,)  \right\} \ .
\end{eqnarray}
This result can  be  written  more  formally by  introducing  a scalar
product of the form:
\begin{equation} 
(\Psi|\Phi)_{0} = \int\!\! d^3\vec{r}\, g(r) \Psi^*(\vec{r}\,) \Phi(\vec{r}\,) \ ,  
\label{eq:weight_sp}
\end{equation}
with the weight ${g(r)=1+\alpha_0 \delta(r)}$\ .

\item Particles interacting in the $p$-wave channel only (for example in the case of 
polarized fermions):
\begin{eqnarray}
(\Psi|\Phi)_{0}= &&\!\!\!\!\!\! \lim_{r_0 \to 0} 
\left\{ \int_{r>r_0}\!\!\!\!\!\! d^3\vec{r} \,  
\Psi^*(\vec{r}\,) \Phi(\vec{r}\,) \right. \nonumber\\
&&\!\!\!\!\!\!
+ \left. (\alpha_1 r_0^4-r_0^3 ) \int_{r=r_0}\!\!\!\!\!\! d^2\Omega \,  
\Psi^*(\vec{r}\,) \Phi(\vec{r}\,)  \right\} \ .
\end{eqnarray}
This scalar product has been first  introduced in Ref.~\cite{pwave} in
the  same form as in Eq.(\ref{eq:weight_sp})  with the  weight  
${g(r)= 1+\delta(r) \left[ (\alpha_1 r^2-r)\, .\, \right]}\ $.
\end{itemize}

\subsection{Bound states and their normalizations}

In this section,  we consider the normalization  of the two-body bound
states      of  vanishing   energy     supported  by  the   zero-range
pseudopotential.  We show that using the regularized scalar product is
equivalent  to the method  based  on the analyticy  of  the scattering
amplitude  \cite{Landau}.     This     last method   was     used   in
Ref.~\cite{Petrov3B} in the context  of $s$-wave narrow resonances.  In
this  case  or  also for   $p$-wave  or higher partial  waves  channel
resonance   where  bound states are   even not  square integrable with
respect to the   usual scalar   product  (they  don't belong  to   the
${\mathcal  L}^2$  Hilbert space),    the regularized scalar   product
appears  as the  natural way  to  perform  the  normalization  in  the
configuration  space. Moreover, this  scalar  product allows for a
generalization  of the method  of   Ref.~\cite{Landau} in  the case  of
inhomogeneous situations.

If one  considers  solutions of the  Schr\"{o}dinger  equation with  a
finite  range potential  as functions  of $\vec{r}$ and  of the energy
$E$:  $\Psi(E,\vec{r}\,)$,  then     a bound  state    of   energy  $E_B$
and wave function $\Psi_B=\Psi(E_B,\vec{r}\,)$ corresponds to a case where $\Psi(E,\vec{r}\,)$ is
square integrable. Using the  Schr\"{o}dinger  equation, one can  show
that:
\begin{eqnarray}
\int_{r < r_0 }\!\!\! d^3\vec{r}\, |\Psi_B|^2 &&\!\!\!\!\!\!= - \frac{\hbar^2 r_0^2}{2 \mu} 
\int_{r=r_0} \!\!\!\!\!\!\!\! d^2\Omega\,
\bigl( \Psi^*(E,\vec{r}\,) \partial_r \partial_E \Psi(E,\vec{r}\,)  \nonumber \\
&&  - \partial_r  \Psi^*(E,\vec{r}\,) \partial_E \Psi(E,\vec{r}\,) \bigl)_{E=E_B} \ .
\label{eq:analyticy}
\end{eqnarray}
Using  this  identity permits  to  deduce that  the normalization of a
bound state  is linked to the  residue of  the scattering amplitude at
${E=E_B}$ \cite{Landau}.  Using  Eq.(\ref{eq:analyticy}), the norm  of
the state $\Psi(E,\vec{r}\,)$ can be also written as:
\begin{eqnarray}
\langle \Psi | \Psi \rangle =  \int_{r>R}\!\!\!\!\!\!  d^3\vec{r}\, |\Psi|^2 
- \frac{\hbar^2 R^2}{2 \mu} &&\!\!\!\!\!\!\!\!
\int_{r=R} \!\!\!\!\!\!\!\! d^2\Omega\, \bigl( \Psi^* \partial_r \partial_E \Psi \nonumber \\
&&  - \partial_r  \Psi^* \partial_E \Psi \bigl)  \ .
\label{eq:norm}
\end{eqnarray}
The  link   between Eq.(\ref{eq:norm})  and the  result  given  by the
regularized    scalar     product   appears   by     noticing     that
Eq.(\ref{eq:analyticy}) is exactly  the opposite  of the \emph{r.h.s.} 
of Eq.(\ref{eq:usual_scalar_product}) in the limit ${k\,'\to k}$ and for
$r_0=R$. As a  result, the norm of  a state  in the zero-range  scheme
obtained through Eq.(\ref{eq:regu_sp}) coincides with the formal limit
$R \to 0$ in Eq.(\ref{eq:norm}).

In  the absence of an  external potential, wave  functions of two body
bound states  of energy ${E_B= -\hbar^2 \kappa^2/2\mu}$  in channel  ($l$) are of
the form (for $r>R$):
\begin{equation}
\Psi_B(r,\Omega) = {\mathcal N}_l \, \phi_l(\Omega)\, 
r^l\! \left(\frac{1}{r}\partial_r\right)^l \! \left(\frac{\exp -\kappa r}{r}\right) \quad, 
\label{eq:bound_state}
\end{equation}
where $\phi_l(\Omega)$  is a normalized angular  function. The  expression for
the  bound state  energy is  found  by solving ${1/f_l(i\kappa)=0}$. For  a
partial scattering amplitude given by Eq.(\ref{eq:partial_amplitude}),
one finds  in the resonant regime \emph{i.e.}   for large and positive
values of $w_l$, a bound state  of vanishing energy with ${\kappa^2 \simeq 1/w_l
\alpha_l}$  (obviously in the case  of the $s$-wave broad resonance $\alpha_0=0$
and ${\kappa=1/w_0}$). This is the state which  plays a crucial role in the
BCS-BEC      crossover.        In     the      zero-range    approach,
Eq.(\ref{eq:bound_state}) is formally extended in  the domain ${0\leq r \leq
R}$ and using  the regularized scalar product,  one finds that for  $(
\Psi_B | \Psi_B )_0 = 1$:
\begin{equation}
{\mathcal N}_l^2 = \frac{1}{\alpha_l + (-1)^l (l+\frac{1}{2}) \kappa^{(2l-1)}} \quad  .
\label{eq:normalization}
\end{equation}
This result is very  interesting as it coincides  with the one given by
the residue of $f_l$ at the energy ${E=E_B}$ \cite{Landau}.

To  conclude  this    section,   we  show  that  the   scalar  product
(\ref{eq:regu_sp}) gives a constraint   on the possible values  of the
``effective range''   parameter $\alpha_l$  compatible with    a consistent
description of the  low-energy bound state and  hence  of the resonant
regime. Indeed, the probability of finding the molecular state outside
the  potential  range  is less  than  one: ${\int_{r>R}  \!  d^3\vec{r}\,
|\Psi_B|^2 <  1}$.   Then, using Eq.(\ref{eq:normalization}) and  keeping
the  dominant contribution  as $R \to  0$, one  finds that the parameter
$\alpha_l$ verifies in the resonant regime for $l>0$:
\begin{equation}
\alpha_l R^{2l-1} \gtrsim (2l-3)!! (2l-1)!! \quad .
\label{eq:condition}
\end{equation}
In   the  case of  a narrow   $s$-wave  resonance,   in the  regime of
intermediate detuning (${\alpha_0\gg w_0}$)  and  ${w_0 \gtrsim R}$,  then  ${\kappa^2 \simeq
1/w_0\alpha_0}$ and the condition ${\alpha_0\gg R}$ follows  directly from ${\kappa R \ll
1}$. By the way, Eq.(\ref{eq:condition}) implies also that ${\alpha_l >0}$.
As an example, the experimental fit performed  in \cite{Ticknor} for a
particular     $p$-wave     resonance    is      compatible       with
Eq.(\ref{eq:condition}).  Let  us  note  that the  ideal   case for  a
description    of the  bound  state    within the zero-range  approach
corresponds   to   the    situation  where     the  two  terms      in
Eq.(\ref{eq:condition})  are  almost  equal.    In this  regime  the
probability  of  finding the dimer in  the  region $r>R$, \emph{i.e.} 
in the region described by the zero-range model, is maximum.

\section{Conclusions}

In  this paper, we   have developed a  general zero-range  one channel
model  for resonant  scattering in arbitrary   partial waves.  In this
approach, the wave function  is a solution of the free  Schr\"{o}dinger
equation  and  the   interaction is replaced    by energy  independent
boundary conditions  on the wave function  as the distance between the
two interacting particles  goes formally    to  zero.  This way,    we
generalize the Bethe-Peierls approach  which  has been widely used  in
the case of broad $s$-wave resonances. We have shown, how the boundary
conditions can be  implemented in the Schr\"{o}dinger equation through
zero-range  pseudopotentials. In a  given  partial  wave channel,  the
interaction can be  described either in  terms of a boundary condition
on the wave function or with a family of pseudopotentials generated by
an extra degree of freedom (the  parameter $\lambda$).  By construction, the
Schr\"{o}dinger equation is invariant  under a change of the parameter
$\lambda$. This   transformation appears then  as   a  general  symmetry  of
zero-range potential approaches.  The pseudopotentials derived in this
paper are the generalization of  $\lambda$-potential already obtained in the
$s$-wave and $p$-wave  channels  \cite{Vlambda,pwave}. This  class  of
pseudopotentials  can serve as   source  terms in the  Schr\"{o}dinger
equation permits us to search solutions with the Green's functions
method.  While exact treatments  are invariant under  a change  of the
$\lambda$ parameter,  this   property is in    general  no longer   true  in
approximate schemes.  This   gives  in  turn   a way  to  improve  the
approximations made (this idea can be applied in mean field treatments
and has been already used in the Hartree-Fock-Bogoliubov formalism for
the Bose gas  \cite{Vlambda,HFB2D}). For example,  we have  shown that
with a specific   choice of $\lambda$,  the  first order Born  approximation
gives the exact   result for two-body  scattering  states at a  finite
colliding energy.

Using the fact that the scattering  states are not mutually orthogonal
in the  zero-range scheme (except  for  the Fermi pseudopotential), we
have  shown how the notion   of a regularized  scalar product  emerges
naturally.  Normalization of bound states by means of this metrics
gives  in  free space  the  same result  as   the method  based on the
analyticy of  the  scattering  amplitude. As  it   applies directly in
configuration  space, this new tool gives  us  a simple way to perform
the normalization in inhomogeneous situations.

The formalism   presented in  this paper   can be used  to  obtain the
scattering amplitude of particles  confined in linear or planar atomic
waveguides, thus  allowing for studies of  ultracold gases trapped in
low dimensions. The few-body problem  is another direction of research
where this approach should be fruitful.

Note added: Recently we learned of  recent related
works  by  Idziaszek  and   Calarco  \cite{Idziaszek} and   also
Derevianko \cite{Derevianko} where the  form of the pseudopotential
in Eq.(\ref{eq:pseudo_stock})  is also found.   The main difference of
the present  approach  with respect to   Ref.~\cite{Idziaszek} (or also
Ref.~\cite{Stock}) is that the pseudopotential in 
Eq.(\ref{eq:pseudopotential}) or equivalently the generalized 
Bethe-Peierls conditions in Eq.(\ref{eq:boundary_general}) have been 
derived in an energy independent form.

\section*{ACKNOWLEDGMENTS}

Y.~Castin,  F.~Chevy, M.~Holzmann  and F.~Werner  are acknowledged for
thorough discussions   on   the  subject.   Laboratoire de    Physique
Th\'{e}orique de  la Mati\`{e}re Condens\'{e}e  is  Unit\'{e} Mixte de
Recherche 7600 of Centre National de la Recherche Scientifique.

\appendix

\section*{APPENDIX: Useful properties of the  symmetric trace free tensors ${\mathcal S}_l$}

This appendix  collects some useful  properties of the symmetric trace
free tensors ${\mathcal      S}_l$ defined in     Eq.(\ref{eq:STF_Sl})
\cite{Courant,Thorne}.  Let  us consider the  two vectors $\vec{r}=x^\alpha
\vec{e}_\alpha$  and    $\vec{r}\,'=x\,'^{\alpha}  \vec{e}_\alpha$. By definition:
\begin{equation}
\frac{1}{|\vec{r}-\vec{r}\,'|} = \sum_{l=0}^\infty \frac{(2l\!-\!1)!!}{l!} {\mathcal S}_{l[\alpha\beta\dots]}  
\frac{x\,'^\alpha x\,'^\beta \dots}{r^{l+1}} \quad .
\label{eq:taylor1}
\end{equation}
Introducing   the  unit  vectors  $\vec{n}=\vec{r}/r=n^\alpha   \vec{e}_\alpha$,
$\vec{n}\,'=\vec{r}\,'/r\,'=n\,'^\alpha \vec{e}_\alpha$ and the symmetric  trace
free tensor ${\mathcal S}_l'$ associated with $\vec{r}\,'$, we have:
\begin{equation}
{\mathcal S}_{l [\alpha\beta\dots]}  n\,'^\alpha n\,'^\beta \dots = 
{\mathcal S}\,'_{l [\alpha\beta\dots]} {\mathcal S}_l^{[\alpha\beta\dots]}  \quad ,
\end{equation}
so that Eq.(\ref{eq:taylor1}) can be rewritten
\begin{equation}
\frac{1}{|\vec{r}-\vec{r}\,'|} = \frac{1}{r} \sum_{l=0}^\infty 
{\mathcal S}_{l [\alpha\beta\dots]}  {\mathcal S}\,'^{[\alpha\beta\dots]}_l  
\frac{(2l\!-\!1)!!}{l!} \left(\frac{r\,'}{r}\right)^l \, .
\label{eq:taylor2}
\end{equation}
This expression can be identified with the standard expansion (for $r\,'<r$) \cite{Morse}:
\begin{equation}
\frac{1}{|\vec{r}-\vec{r}\,'|} =\frac{1}{r} \sum_{l=0}^{\infty} 
\left(\frac{r\,'}{r}\right)^l 
{\mathcal P}_l(\vec{n}.\vec{n}\,') \quad ,
\label{eq:taylor_Legendre}
\end{equation}
where ${\mathcal P}_l(x)$ is the Legendre polynomial of degree $l$ and we obtain:
\begin{equation}
{\mathcal P}_l(\vec{n}.\vec{n}\,') = \frac{(2l\!-\!1)!!}{l!}
{\mathcal S}_{l [\alpha\beta\dots]} 
{\mathcal S}\,'^{[\alpha\beta\dots]}_l \quad .
\label{eq:Pl_Sl}
\end{equation}
For example, taking $\vec{n}\,'=\vec{e}_z$ shows that:
\begin{equation}
{\mathcal S}_{l (z z z \dots z)} = \frac{l!}{(2l\!-\!1)!!} 
{\mathcal P}_l(\cos \theta)
\quad ,
\end{equation}
with  $\theta=(\widehat{\vec{r},\vec{e}_z})$. Eq.(\ref{eq:Pl_Sl})  can   be
used to have an expression  of the projection  operator $\Pi_l$ over the
partial  wave $l$ in  terms of the tensor  ${\mathcal  S}_l$. For that
purpose, we recall the addition theorem
\cite{Morse}:
\begin{equation}
{\mathcal P}_l(\vec{n}.\vec{n}\,' ) = \frac{4\pi}{2l\!+\!1} 
\sum_{m=-l}^{m=l} Y^{lm}(\Omega) Y^{lm\,*}(\Omega\,') \quad ,
\end{equation}
where  $(\Omega=(\theta,\phi)$,  $\Omega\,'=(\theta',\phi')$)
are the angles  defining  the spherical coordinates for  
the two vectors ($\vec{r},\vec{r}\,'$). The projection of a wave 
function $\Psi$ reads:
\begin{equation}
\langle \vec{r}\,| \Pi_l | \Psi \rangle = 
(2l\!+\!1) \int\!\! \frac{d^2 \Omega\,'}{4 \pi} \Psi(r,\Omega\,')  
{\mathcal P}_l(\vec{n}.\vec{n}\,' ) \quad .
\label{eq:projector_Pl}
\end{equation}
together with Eq.(\ref{eq:Pl_Sl}) and the definition of the multipoles
of     the   wave  function    Eq.(\ref{eq:multipolar}),   we   obtain
Eq.(\ref{eq:projector}).

In  this paper, we also use  an interesting property  of the symmetric
traceless tensors found    in Ref.~\cite{Thorne}. Let  us  consider two
symmetric and  traceless tensors ${\mathcal A}$  and ${\mathcal B}$ of
order $l$, then
\begin{eqnarray}
{\mathcal A}^{[\alpha\beta\dots]}  \left( \int \! \frac{d^2\Omega\,'}{4\pi}
n\,'_\alpha n\,'_\beta \dots n\,'_{\alpha'} n\,'_{\beta'} \right)
&&\!\!\!\!\!\! {\mathcal B}^{[\alpha'\beta'\dots]} \nonumber\\
= \frac{l!}{(2l\!+\!1)!!} 
{\mathcal A}_{[\alpha\beta\dots]}
&&\!\!\!\!\!\! {\mathcal B}^{[\alpha\beta\dots]} \quad .
\label{eq:thorne}
\end{eqnarray}
As an example,  taking ${\mathcal  A}={\mathcal S}_l$ and   ${\mathcal
B}={\mathcal  S}\,''_l$ allows one to recover from
Eqs.(\ref{eq:Pl_Sl},\ref{eq:thorne}) the known relation:
\begin{equation}
{\mathcal P}_l(\vec{n}.\vec{n}\,'') = 
(2l\!+\!1)\! \int \!\! \frac{d^2\Omega\,'}{4\pi} 
 {\mathcal P}_l(\vec{n}.\vec{n}\,') {\mathcal P}_l(\vec{n}\,'.\vec{n}\,'') .
\end{equation}

We close this  appendix with the link  between the singularity of  the
$l$th partial wave of  a wave function  with  the derivatives  of the
delta distribution.  To this end, we use the so-called relation:
\begin{equation}
\Delta_{\vec{r}}\left(\frac{1}{|\vec{r}-\vec{r}\,'|}\right) 
= - 4\pi \delta(\vec{r}-\vec{r}\,'\,)
\end{equation}
together with Eq.(\ref{eq:taylor2}) and the formal expansion:
\begin{equation}
\delta(\vec{r}-\vec{r}\,'\,)  = \sum_{l=0}^\infty  \frac{(-r\,')^l}{l!} 
(\partial\,^l_{[\alpha\beta\dots]} \delta)(\vec{r}\,) n\,'^\alpha n\,'^\beta \dots \ ,
\end{equation}
where  $(\partial\,^l_{[\alpha\beta\dots]} \delta)(\vec{r}\,)$   is      the  $l$th  partial
derivative  of the  delta distribution (see Eq.(\ref{eq:partial})). Finally, we obtain:
\begin{eqnarray}
&&\!\!\!\!\!\!\!\! \Delta_{\vec{r}} \left( \frac{{\mathcal S}_{l [\alpha\beta\dots]}}{r^{l+1}}\right) 
n\,'^\alpha n\,'^\beta \dots = \nonumber \\
&& \qquad 4\pi \frac{(-1)^{l+1}}{(2l\!-\!1)!!} (\partial\,^l_{[\alpha\beta\dots]} \delta)(\vec{r}\,) 
n\,'^\alpha n\,'^\beta \dots  \quad .
\label{eq:singularity}
\end{eqnarray}
This   expression  justifies   the  introduction  of   the   potential
$V_l^{(i)}$        in     Eq.(\ref{eq:pseudo(i)}).      Using
Eq.(\ref{eq:Pl_Sl}),  one  can also write Eq.(\ref{eq:singularity}) in
terms of Legendre Polynomials:
\begin{equation} 
\Delta_{\vec{r}}\left( \frac{{\mathcal P}_l(\cos \theta)}{r^{l+1}} \right) = 
4\pi \frac{(-1)^{l+1}}{l!} (\partial_z^{\,l} \delta)(\vec{r}\,) \quad .
\label{eq:singular_legendre}
\end{equation}


\begin{thebibliography}{99}


\bibitem{Bethe} H. Bethe and R. Peierls, Proc. R. Soc. London, Ser. 
A {\bf 148}, 146 (1935).

\bibitem{Fermi} E.  Fermi, Ric. Sci. {\bf 7-II}, 13 (1936). 

\bibitem{Breit} G. Breit, Phys. Rev. {\bf 71}, 215 (1947). 

\bibitem{Blatt} J.M. Blatt and V.F. Weisskopf, in 
\textit{Theoretical Nuclear Physics}, Wiley, New York (1952). 


\bibitem{Cohen} C. Cohen-Tannoudji, Course at Coll\`{e}ge de France, Lectures 4,5 (1998-99).
http://www.phys.ens.fr/cours/college-de-france/1998-99/1998-99.htm

\bibitem{YvanLesHouches} Y. Castin, in \textit{Coherent Atomic Matter Waves},
Lecture Notes of Les Houches  Summer School, EDP Sciences and
(Springer-Verlag, Berlin, 2001), pp.1-136.

\bibitem{BruunBCS}  G. Bruun, Y. Castin, R. Dum and K. Burnett, Eur. Phys. J. D {\bf 7}, 433 (1999).

\bibitem{Vlambda} M. Olshanii and L. Pricoupenko, Phys. Rev. Lett. {\bf 88}, 010402 (2002).


\bibitem{HFB2D} L. Pricoupenko, Phys. Rev. A {\bf 70}, 013601 (2004).

\bibitem{QGLDPetrov} D.S. Petrov, D.M. Gangardt and G.V.~Shlyapnikov, J. Phys. IV France {\bf 64}, 
5 (2004), and references therein.

\bibitem{QGLDMaxim} M.G. Moore, T. Bergeman and M. Olshanii, J. Phys. IV France {\bf 116}, 69
(2004), and references therein.

\bibitem{Peano}  V. Peano, M. Thorwart, C. Mora, and R. Egger, New J. Phys. {\bf 7}, 192 (2005).



\bibitem{Petrov3F} D.S. Petrov, Phys. Rev. A {\bf 67}, 010703(R) (2003).  
 
\bibitem{Petrovscatt} D.S. Petrov, C. Salomon, and G.V.~Shlyapnikov, 
Phys. Rev. A {\bf 71}, 012708(R) (2005).

\bibitem{Petrov3B} D.S. Petrov, Phys. Rev. Lett. {\bf 93}, 143201 (2004). 

\bibitem{Mora} C. Mora, R. Egger, A.O. Gogolin, and A. Komnik,  Phys. Rev. Lett. {\bf 93}, 170403 (2004).


\bibitem{Ketterle_Feshbach} S.~Inouye, M.R.~Andrews, J.~Stenger, H.-J.~Miesner,  D.M.~Stamper-Kurn, 
and W.~Ketterle, Nature (London) {\bf 392}, 151 (1998).
 

\bibitem{Jochim} S. Jochim, M.~Bartenstein, A.~Altmeyer, G.~Hendl, S.~Riedl, C.~Chin, J.~Hecker Denschlag, 
and R.~Grimm, Science {\bf 302}, 2101 (2003).

\bibitem{Zwierlein} M.W.~Zwierlein, C.A.~Stan, C.H.~Schunck, S.M.F.~Raupach, S.~Gupta, Z.~Hadzibabic, and 
W.~Ketterle, Phys. Rev. Lett. {\bf 91}, 250401 (2003).

\bibitem{Bourdel} T.~Bourdel, L.~Khaykovich, J.~Cubizolles, J.~Zhang, F.~Chevy, M.~Teichmann, L.~Tarruell, 
S.J.J.M.F.~Kokkelmans, and C.~Salomon, Phys. Rev. Lett. {\bf 93}, 050401 (2004). 

\bibitem{Kinast} J.~Kinast, S.L.~Hemmer, M.E.~Gehm, A.~Turlapov, and J.E.~Thomas, Phys. Rev. Lett. 
{\bf 92}, 150402 (2004).

\bibitem{Greiner} Markus~Greiner, Cindy~A.~Regal, and Deborah~S.~Jin, Nature (London) {\bf 426}, 537 (2003).


\bibitem{Yvanscaling} Y. Castin,  C. R. Phys. {\bf 5}, 407 (2004).

\bibitem{Felix_OH} F. Werner and Y. Castin, \textit{cond-mat/0507399}.


\bibitem{unitary} M.W. Zwierlein, J.R. Abo-Shaeer, A. Schirotzek, C.H. Schunck, and W. Ketterle, Nature 
(London) {\bf 435}, 1047 (2005).

\bibitem{NSR} P. Nozi\`{e}res and S. Schmitt-Rink, J. Low Temp. Phys. {\bf 59}, 195 (1985).

\bibitem{Randeria} M. Randeria, in {\it Bose-Einstein Condensation},
edited by A. Griffin, D. W. Snoke, S. Stringari (Cambridge University Press,
Cambridge, England, 1995), p.355.


\bibitem{Regal} C.A. Regal, C. Ticknor, J.L. Bohn, and D.S. Jin, Phys. Rev. Lett. {\bf 90}, 053201 (2003). 

\bibitem{Ticknor}  C. Ticknor, C.A. Regal, D.S.~Jin, and J.L.~Bohn, Phys. Rev. A {\bf 69}, 042712 (2004).

\bibitem{Zhang} J. Zhang, E.G.M. van Kempen, T. Bourdel, L. Khaykovich, J.~Cubizolles, F.~Chevy, 
M.~Teichmann, L.~Tarruell, S.J.J.M.F.~Kokkelmans, and C.~Salomon, Phys. Rev. A {\bf 70}, 030702(R) (2004).

\bibitem{Schunck} C.H. Schunck, M.W. Zwierlein, C.A. Stan, S.M.F.~Raupach, W.~Ketterle, 
A.~Simoni, E.~Tiesinga, C.J.~Williams, and P.S.~Julienne, Phys. Rev. A  {\bf 71}, 045601 (2005).

\bibitem{Chevy} F. Chevy, E.G.M. Van Kempen, T.~Bourdel, J.~Zhang, L.~Khaykovich, 
M.~Teichmann, L.~Tarruell, S.J.J.M.F.~Kokkelmans, and C.~Salomon, Phys. Rev. A {\bf 71}, 062710 (2005).
 

\bibitem{Volz_1} T. Volz, S. D\"{u}rr, N. Syassen, G. Rempe, E.G.M.~van~Kempen, and S.J.J.M.F.~Kokkelmans, 
Phys. Rev. A {\bf 72}, 010704(R) (2005).


\bibitem{Stenger} J. Stenger, S. Inouye, M.R. Andrews, H.-J.~Miesner, D.M.~Stamper-Kurn, 
and W.~Ketterle, Phys. Rev. Lett. {\bf 82}, 2422 (1999).



\bibitem{pwave} L. Pricoupenko, {\sl cond-mat/0505448}.


\bibitem{Huang} K. Huang and C. N. Yang, Phys. Rev. {\bf 105}, 767 (1956).

\bibitem{Stock} R. Stock,~A. Silberfarb,~E.L.~Bolda and~I.H.~Deutsch, Phys. Rev. Lett. {\bf 94}, 023202 (2005). 


\bibitem{Landau} L. Landau et E. Lifchitz, Tome III \textit{M\'{e}canique 
Quantique}, Editions MIR Moscou (1967).

\bibitem{Dalibard} J.~Dalibard, in \textit{Proceedings of the International 
School of   Physics   ``Enrico Fermi''}, edited by M.~Inguscio,
S.~Stringari and C.~Wieman,  Course   CXL (IOS, Amsterdam, 1999), pp.321-349.

\bibitem{Marinescu} M. Marinescu, H.R. Sadeghpour and A.~Dalgarno, 
Phys. Rev. A {\bf 49}, 982 (1994).

\bibitem{Mott} N.F. Mott and M.S. Massey, in \textit{The Theory of Atomic 
Collisions}, $3$rd ed. (Clarendon, Oxford, 1965).

\bibitem{Feshbach} H. Feshbach, Ann. Phys. (N.Y.) {\bf 19}, 287 (1962).


\bibitem{Abramowitz} M. Abramowitz and Irene Stegun, in \textit{Handbook of 
Mathematical Functions} (Dover Publications, Inc., New York, 1974).

\bibitem{Courant} R. Courant and D. Hilbert, in \textit{Methods of Mathematical 
Physics} Vol. I (John Wiley, New-York,1989).


\bibitem{Roth} R. Roth and H. Feldmeier, Phys. Rev. A {\bf 64}, 043603 (2001).

\bibitem{Box} L. Pricoupenko and Y. Castin, Phys. Rev. A {\bf 69}, 051601(R) (2004). 


\bibitem{Idziaszek} Z.~Idziaszek and T.~Calarco, {\it quant-ph/0507186}.

\bibitem{Derevianko} A.~Derevianko, Phys. Rev. A {\bf 72}, 044701 (2005).


\bibitem{Thorne} K.S. Thorne, Rev. Mod. Phys. {\bf 52}, No 2, Part I, 299 (1980).

\bibitem{Morse} P.M. Morse and H. Feshbach, in \textit{Methods of Theoretical Physics} (Mc Graw-Hill, New-York, 1953). 


\end{thebibliography}
\end{document}